\documentclass{article}

\usepackage[french]{babel}
\usepackage{geometry}    %permet la mise en page
\usepackage[latin1]{inputenc} %permet de reconnaître les accents
\usepackage[T1]{fontenc} %permet de reconnaître les accents
\usepackage{amsfonts}    %permet d'écrire des lettres doublées (comme N, Z, D, Q, R, C, H)
\usepackage{amssymb}     %permet d'écrire des symboles comme \simeq, \cong, \bowtie ATTENTION ALORS AUX MATRICES
\usepackage[all]{xy}     %permet de tracer des diagrammes
\usepackage{amsmath}     %permet de superposer des indices dans une somme avec \substack
\usepackage{stmaryrd}    %permet d'écrire des intervalles d'entiers avec \llbracket

\newtheorem{The}{Théorème}
\newtheorem{Def}[The]{Définition}

\newtheorem{Pro}[The]{Propriété}

\newtheorem{Rq}[The]{Remarque}

\begin{document}

\begin{center}
\Large{COHOMOLOGIE DE HOCHSCHILD DES SURFACES DE
KLEIN}\vspace{.2cm}\\
\end{center}

\begin{center}
Frédéric BUTIN\footnote{butin@math.univ-lyon1.fr}\vspace{.2cm}\\
\end{center}

\begin{small}
\noindent\textbf{\textsc{Résumé}}\\
Etant donné un système physique $(M,\,\mathcal{F}(M))$, où $M$ est
une variété de Poisson et $\mathcal{F}(M)$ l'algèbre des fonctions
régulières sur $M$, il est important de pouvoir le quantifier pour
obtenir des résultats plus corrects que ceux donnés par la
mécanique classique. Une solution est fournie par la
quantification par déformation qui consiste à construire un
star-produit sur l'algèbre des séries formelles
$\mathcal{F}(M)[[\hbar]]$. Un premier pas vers l'étude des
star-produits est le calcul de la cohomologie de Hochschild de
$\mathcal{F}(M)$.\\
Le but de l'article est de déterminer cette cohomologie de
Hochschild dans le cas des courbes singulières du plane --- on
précise ainsi, par une démarche différente, un résultat démontré
par Fronsdal --- et dans le cas des surfaces de Klein.
L'utilisation d'un complexe proposé par Kontsevich et l'emploi des
bases de Gröbner permettent de résoudre le problème.\\
\end{small}

\begin{small}
\noindent\textbf{\textsc{Abstract}}\\
Given a mechanical system $(M,\,\mathcal{F}(M))$, where $M$ is a
Poisson manifold and $\mathcal{F}(M)$ the algebra of regular
functions on $M$, it is important to be able to quantize it, in
order to obtain more precise results than through classical
mechanics. An available method is the deformation quantization,
which consists in constructing a star-product on the algebra of
formal power series $\mathcal{F}(M)[[\hbar]]$. A first step toward
study of star-products is the calculation of Hochschild cohomology
of
$\mathcal{F}(M)$.\\
The aim of this article is to determine this Hochschild cohomology
 in the case of singular curves of the plane --- so we
rediscover, by a different way, a result proved by Fronsdal and
make it more precise --- and in the case of Klein surfaces. The
use of a complex suggested
 by Kontsevich and the help of Gröbner bases allow us to solve the problem.\\
\end{small}

\begin{small}
\noindent\textbf{\textsc{Mots-clés}}\\
cohomologie ; Hochschild ; surfaces de Klein ; bases de Gröbner ;
quantification ; star-produits.
\end{small}

\tableofcontents

\section{\textsf{Introduction}}

\subsection{\textsf{Quantification par déformation}}

\noindent On considère un système physique donné par une variété
de Poisson $M$, munie du crochet de Poisson $\{\cdot\}$. En
mécanique classique, on étudie l'algèbre (commutative)
$\mathcal{F}(M)$ des fonctions régulières (ie, par exemple,
$\mathcal{C}^\infty$, holomorphes ou polynomiales) sur $M$,
c'est-à-dire des observables de la mécanique classique. Or la
mécanique quantique, où le système physique est décrit par une
algèbre (non commutative) d'opérateurs sur un espace de Hilbert,
donne des résultats plus corrects que son analogue classique. D'où
l'intérêt d'obtenir une description quantique du système classique
$(M,\,\mathcal{F}(M))$$\ $ : une telle opération s'appelle une
quantification. Une des solutions est la quantification
géométrique qui permet de construire explicitement un espace de
Hilbert et une algèbre d'opérateurs sur cet espace. Cette méthode,
fort intéressante, a l'inconvénient de ne pas toujours
s'appliquer. C'est pourquoi on a introduit d'autres
quantifications telles la quantification asymptotique et la
quantification par déformation. Cette dernière, décrite en 1978
par F. Bayen, M. Flato, C. Fronsdal, A. Lichnerowicz et D.
Sternheimer dans l'article \cite{BFFLS78}, constitue une bonne
alternative : au lieu de construire une algèbre d'opérateurs sur
un espace de Hilbert, il s'agit d'obtenir une déformation formelle
de $\mathcal{F}(M)$, donnée par l'algèbre des séries formelles
$\mathcal{F}(M)[[\hbar]]$, munie du star-produit associatif (mais
non commutatif)
\begin{equation}\label{def}
  f\ast
g=\sum_{j=0}^\infty m_j(f,\,g)\hbar^j
\end{equation}
où les applications $m_j$ sont bilinéaires et où $m_0(f,\,g)=fg$.
La quantification est alors donnée par l'application $f\mapsto
\widehat{f}$, où $\widehat{f}(g)=f\ast g$.
\\ On peut se demander dans
quels cas une variété de Poisson admet une telle quantification.
Une réponse a été donnée par Kontsevich dans son article
\cite{K97} : il a en effet construit un star-produit sur toute
variété de Poisson. En outre, il a démontré que si $M$ est une
variété lisse, alors les classes d'équivalence de déformations
formelles du crochet de Poisson nul sont en bijection avec les
classes d'équivalence de star-produits. De plus, d'après le théorème de Hochschild-Kostant-Rosenberg,
tout star-produit abélien est trivial.\\
Dans le cas où $M$ est une variété algébrique singulière, de la
forme $$M=\{\mathbf{z}\in \mathbb{C}^n\ /\ f(\mathbf{z})=0\}$$
avec $n=2$ ou $3$, où $f$ est un polynôme de
$\mathbb{C}[\mathbf{z}]$ --- et c'est le cas que nous étudions
dans la suite --- les fonctions régulières à considérer sont les
fonctions polynomiales sur $M$, dont l'algèbre s'identifie à
l'algèbre quotient $\mathbb{C}[\mathbf{z}]\,/\,\langle f\rangle$.
Le résultat énoncé précédemment n'est donc plus applicable.
Cependant, les déformations de l'algèbre $\mathcal{F}(M)$,
définies par la formule~(\ref{def}), sont toujours classifiées par
la cohomologie de Hochschild de $\mathcal{F}(M)$, et on se trouve
ainsi ramené à l'étude de la cohomologie de Hochschild de
$\mathbb{C}[\mathbf{z}]\,/\,\langle
f\rangle$.\\

\subsection{\textsf{Cohomologies et quotients d'algèbres de polynômes}}

\noindent Dans la suite, on considère $R=\mathbb{C}[z_1,\ \dots,\
z_n]=\mathbb{C}[\mathbf{z}]$ l'algèbre des polynômes à
coefficients complexes et à $n$ indéterminées. On fixe aussi
$f_1,\ \dots,\ f_m$ $m$ éléments de $R$, et on définit l'algèbre
quotient $A:=R\ /\ \langle f_1,\ \dots,\
f_m\rangle=\mathbb{C}[z_1,\ \dots,\ z_n]\ /\
\langle f_1,\ \dots,\ f_m\rangle$.\\

\noindent Plusieurs articles ont été consacrés à l'étude de cas
particuliers :\\

\begin{minipage}{15.4cm}
\noindent C. Roger et P. Vanhaecke, dans l'article \cite{RV02},
considèrent le cas où $n=2$ et $m=1$, et où $f_1$ est un polynôme
homogène. Après avoir rappelé la définition de la cohomologie de
Poisson, ils la calculent en fonction du nombre de composantes
irréductibles du lieu singulier $\{\mathbf{z}\in \mathbb{C}^2\ /\
f_1(\mathbf{z})=0\}$ (dans ce cas, on a une structure symplectique
en dehors du lieu singulier).\\

\noindent M. Van den Bergh et A. Pichereau, dans les articles
\cite{VB94}, \cite{P05} et \cite{P06}, s'intéressent au cas où
$n=3$ et $m=1$, et où $f_1$ est un polynôme quasi-homogène à
singularité isolée en l'origine. Ils présentent le calcul de
l'homologie et de la cohomologie de Poisson, qui s'exprime en
particulier en fonction du nombre de Milnor de l'espace
$\mathbb{C}[z_1,\,z_2,\,z_3]\ /\ \langle
\partial_{z_1}f_1,\,\partial_{z_2}f_1,\,\partial_{z_3}f_1\rangle$.\\

\noindent En s'intéressant toujours au cas où $n=3$ et $m=1$, dans
l'article \cite{AL98}, J. Alev et T. Lambre comparent l'homologie
de Poisson en degré $0$ des surfaces de Klein à l'homologie de
Hochschild en degré $0$ de $A_1(\mathbb{C})^G$, où
$A_1(\mathbb{C})$ est l'algèbre
de Weyl et $G$ le groupe associé à la surface.\\

\noindent Quant à C. Fronsdal, il étudie dans l'article
\cite{FK07} l'homologie et la cohomologie de Hochschild dans deux
cas particuliers : le cas où $n=1$ et $m=1$, et le cas où $n=2$ et
$m=1$. De plus, l'appendice de cet article donne un autre moyen de
calculer la
cohomologie de Hochschild dans le cas plus général des intersections complètes.\\
\end{minipage}

\noindent Dans cet article, on se propose de calculer la
cohomologie de Hochschild dans deux cas particulièrement
intéressants : le cas des courbes singulières du plan, avec des
polynômes $f_1$ qui correspondent à leurs formes normales (ce cas
a déjà retenu l'intérêt de C. Fronsdal) ; et le cas des surfaces
de Klein ($n=3$ et~$m=1$). Ces dernières ont fait l'objet d'un
nombre important de travaux ; leur rapport aux sous-groupes finis
de $\mathbf{SL}_2\mathbb{C}$, aux solides de Platon, et à la
correspondance de McKay explique cet intérêt manifeste. Par
ailleurs, les algèbres préprojectives, dont il est question dans
l'article \cite{CBH98}, constituent une famille de déformations
des surfaces de Klein, paramétrée par le groupe qui leur est
associé, ce qui motive encore le calcul de leur cohomologie.\\
Le résultat principal de l'article est donné par les deux propriétés :\\

\begin{Pro}$\\$
Soit une courbe singulière du plan, définie par un polynôme
$f_1\in\mathbb{C}[\mathbf{z}]$, de type $A_k$, $D_k$ ou $E_k$.
Alors $H^0\simeq \mathbb{C}[\mathbf{z}]\,/\,\langle f_1\rangle$,
$H^1\simeq \mathbb{C}[\mathbf{z}]\,/\,\langle
f_1\rangle\,\oplus\,\mathbb{C}^k$, et pour tout $j\geq 2$, $H^j\simeq\mathbb{C}^k$.\\
\end{Pro}

\begin{Pro}$\\$
Soient $\Gamma$ un sous-groupe fini de $\mathbf{SL}_2\mathbb{C}$
et $f_1\in\mathbb{C}[\mathbf{z}]$ tel que
$\mathbb{C}[x,\,y]^\Gamma\simeq\mathbb{C}[\mathbf{z}]\,/\,\langle
f_1\rangle$. Alors $H^0\simeq \mathbb{C}[\mathbf{z}]\,/\,\langle
f_1\rangle$, $H^1\simeq \nabla f_1
\wedge(\mathbb{C}[\mathbf{z}]\,/\,\langle
f_1\rangle)^3\,\oplus\,\mathbb{C}^\mu$, $H^2\simeq
\mathbb{C}[\mathbf{z}]\,/\,\langle
f_1\rangle\,\oplus\,\mathbb{C}^\mu$, et pour tout $j\geq 3$,
$H^j\simeq\mathbb{C}^\mu$, où $\mu$ est le nombre de Milnor
de $\mathcal{X}_\Gamma$.\\
\end{Pro}

\noindent Pour calculer explicitement ces espaces de cohomologie,
on s'appuiera sur la démarche proposée par M. Kontsevich dans
l'appendice de \cite{FK07}, démarche que l'on développera.\\
On étudiera d'abord le cas des courbes singulières du plan dans le
paragraphe 3 : on utilisera cette méthode pour retrouver le
résultat que C.~Fronsdal a établi par des calculs directs. Puis on
l'affinera en déterminant les dimensions des espaces de
cohomologie au moyen de
la division multivariée et des bases de Gröbner.\\
Puis, dans le paragraphe 4, on considérera le cas des surfaces de
Klein ($\mathcal{X}_\Gamma=\mathbb{C}^2\,/\,\Gamma$, avec $\Gamma$
sous-groupe fini de $\mathbf{SL}_2\mathbb{C}$). On démontrera
d'abord que $H^0$ s'identifie à l'espace des fonctions
polynomiales sur la surface singulière~$\mathcal{X}_\Gamma$. On
poursuivra en montrant que $H^1$ et $H^2$ sont de dimension
infinie. On déterminera aussi, pour $j$ supérieur ou égal à~$3$,
la dimension de $H^j$, en montrant qu'elle est égale au nombre de
Milnor de la surface~$\mathcal{X}_\Gamma$.\\
\noindent Avant l'étude de ces deux cas, le paragraphe 1.3
rappelle des
résultats importants sur les déformations.\\

\subsection{\textsf{Cohomologie de Hochschild et déformations d'algèbres}}

\noindent $\bullet$ Etant donné une $\mathbb{C}-$algèbre
associative, notée $A$, le complexe de Hochschild associé à $A$
est le complexe
$$\xymatrix{C^0(A) \ar@{->}[r]^{d_0} &
C^1(A) \ar@{->}[r]^{d_1}
 & C^2(A) \ar@{->}[r]^{d_2}
 & C^3(A) \ar@{->}[r]^{d_3}
 & C^4(A) \ar@{->}[r]^{d_4}
 & \dots }$$
dont l'espace $C^p(A)$ des $p-$cochaînes est défini par $C^p(A)=0$
pour $p\in -\mathbb{N}^*$, $C^0(A)=A$ et $\forall\ p\in
\mathbb{N}^*,\ C^p(A)=L(A^{\otimes p},\ A)$, où $L(A^{\otimes p},\
A)$ désigne l'espace des applications $\mathbb{C}-$linéaires de
$A^{\otimes p}$ dans $A$, et dont la différentielle
$d=\bigoplus_{i=0}^\infty d_p$ est donnée par la formule
$$\forall\ f\in C^p(A),\ d_p\,f(a_0,\dots,\,a_p)=a_0f(a_1,\dots,\,a_p)-\sum_{i=0}^{p-1}f(a_0,\dots,\,a_ia_{i+1},\dots,\,a_p)
+(-1)^{p-1}f(a_0,\dots,\,a_{p-1})a_p,$$
$$\textrm{c'est-à-dire}\ d_pf=(-1)^{p+1}[\mu,\,f]_G,$$
où $\mu$ est la multiplication de l'algèbre $A$, et $[\cdot]_G$ le
crochet de Gerstenhaber.\\
On définit alors la cohomologie de
Hochschild de $A$ comme la cohomologie du complexe de Hochschild
associé à $A$. On note $HH^0(A)=\textrm{Ker}\ d_0$ et $\forall\ p\in \mathbb{N}^*,\ HH^p(A)=\textrm{Ker}\ d_p\ /\ \textrm{Im}\ d_{p-1}$.\\

\noindent $\bullet$ On note $\mathbb{C}[[\hbar]]$ (resp.
$A[[\hbar]]$) l'algèbre des séries formelles en l'indéterminée
$\hbar$, à coefficients dans $\mathbb{C}$ (resp. $A$). Une
déformation de l'algèbre $A$ est définie comme une application $m$
de $A[[\hbar]]\times A[[\hbar]]$ dans $A[[\hbar]]$ qui est
$\mathbb{C}[[\hbar]]-$bilinéaire et telle que $$\begin{array}{l}
\forall\ (s,\,t)\in
A[[\hbar]]^2,\ \ m(s,\,t)\equiv st \mod \hbar A[[\hbar]], \\
\forall\ (s,\,t,\,u)\in
A[[\hbar]]^3,\ \ m(s,\,m(t,\,u))=m(m(s,\,t),\,u).\\
\end{array}$$\\
Cela signifie qu'il existe une suite d'applications bilinéaires
$m_j$ de $A\times A$ dans $A$ dont le premier terme $m_0$ est le
produit de $A$ et telle que
$$\begin{array}{l}
  \displaystyle{\forall\ (a,\,b)\in A^2,\
m(a,\,b)=\sum_{j=0}^{\infty}m_j(a,\,b)\hbar^j}, \\
  \displaystyle{\forall\ n\in \mathbb{N},\ \sum_{i+j=n}m_i(a,\,m_j(b,\,c))=\sum_{i+j=n}m_i(m_j(a,\,b),\,c),\ \textrm{c'est-à-dire}\
  \sum_{i+j=n}m_i\bullet m_j=0,}\\
\end{array}$$
en utilisant le produit de Gerstenhaber, noté $\bullet$.\\
On parle
de déformation d'ordre $p$ si la formule précédente est
vérifiée (seulement) pour $n\leq p$.\\
Deux déformations $m$ et $m'$ de $A$ sont dites équivalentes s'il
existe un $\mathbb{C}[[\hbar]]-$automorphisme de $A[[\hbar]]$,
noté $\varphi$,  tel que
$$\begin{array}{l}
  \forall\ (s,\,t)\in A[[\hbar]]^2,\
\varphi(m(s,\,t))=m'(\varphi(s),\,\varphi(t)) \\
  \forall\ s\in A[\hbar],\ \varphi(s)\equiv s \mod \hbar A[[\hbar]], \\
\end{array}$$\\
c'est-à-dire s'il existe une suite d'applications linéaires
$\varphi_j$ de $A$ dans $A$ dont le premier terme $\varphi_0$ est
l'identité de $A$ et telle que
$$\begin{array}{l}
  \displaystyle{\forall\ a\in A,\
\varphi(a)=\sum_{j=0}^{\infty}\varphi_j(a)\hbar^j}, \\
  \displaystyle{\forall\ n\in \mathbb{N},\ \sum_{i+j=n}\varphi_i(m_j(a,\,b))=\sum_{i+j+k=n}m'_i(\varphi_j(a),\,\varphi_k(b)).}\\
\end{array}$$

\noindent $\bullet$ Un des intérêts de la cohomologie de
Hochschild est de permettre de paramétrer les déformations de
l'algèbre $A$. En effet, si $\pi\in C^2(A)$, on peut construire
une déformation $m$ d'ordre $1$ de $A$ telle que $m_1=\pi$ si et
seulement si $\pi\in \textrm{Ker}\,d_2$. De plus, deux telles
déformations sont équivalentes si et seulement si leur différence
est un élément de $\textrm{Im}\,d_1$. Ainsi, l'ensemble des
classes de déformations d'ordre $1$ est en bijection avec
$HH^2(A)$.\\
Si $m$ est une déformation d'ordre $p$, alors on peut étendre $m$
en une déformation d'ordre $p+1$ si et seulement s'il existe
$m_{p+1}$ tel que
$$\begin{array}{c}
\forall\ (a,\,b,\,c)\in A^3,\ \underbrace{\sum_{i=1}^p\left(m_i(a,\,m_{p+1-i}(b,\,c))-m_i(m_{p+1-i}(a,\,b),\,c)\right)}_{\omega_p(a,\,b,\,c)}=-d_2\,m_{p+1}(a,\,b,\,c), \\
\\
\textrm{ie}\ \displaystyle{\sum_{i=1}^p m_i\bullet m_{p+1-i}=d_2m_{p+1}.} \\
\end{array}$$
Or le terme $\omega_p$ appartient à $\textrm{Ker}\,d_3$, donc
$HH^3(A)$ représente les obstructions au prolongement d'une
déformation d'ordre $p$ en une déformation d'ordre $p+1$.\\

\section{\textsf{Présentation du complexe de Koszul}}

\noindent On rappelle dans ce paragraphe les résultats sur le
complexe de
Koszul qui sont donnés dans l'appendice de l'article \cite{FK07}.\\

\subsection{\textsf{Théorème de Kontsevich et notations}}

\noindent $\bullet$ Comme indiqué au paragraphe 1.2, on considère
$R=\mathbb{C}[\mathbf{z}]$ et $(f_1,\,\dots,\,f_m)\in R^m$, et on
note $A=R\ /\ \langle f_1,\ \dots,\ f_m\rangle$.\ On suppose
 qu'il y a \emph{intersection complète}, ie la dimension de l'ensemble des solutions du système
 $\{f_1=\dots=f_m=0\}$ est $n-m$.\\

\noindent $\bullet$ On définit aussi la superalgèbre
supercommutative
 $\widetilde{A}=R\,\otimes \bigwedge \{\alpha_j,\ j=1\dots m\}=\mathbb{C}[z_1,\ \dots,\ z_n,\ \alpha_1,\ \dots,\
 \alpha_m]$.\\
 On introduit ensuite $\eta_i=\frac{\partial}{\partial z_i}$ et
 $b_j=\frac{\partial}{\partial\alpha_j}$.\\
On note les variables paires avec des lettres latines et
les variables impaires avec des lettres grecques.\\

\noindent $\bullet$ On considère l'algèbre différentielle graduée
$$\widetilde{T}=A[\eta_1,\ \dots,\ \eta_n,\ b_1,\ \dots,\
 b_m]=\frac{\mathbb{C}[z_1,\ \dots,\ z_n]}{\langle f_1,\ \dots,\ f_m\rangle}[\eta_1,\ \dots,\ \eta_n,\ b_1,\ \dots,\
 b_m],$$
munie de la différentielle
$$\fbox{$\displaystyle{d_{\widetilde{T}}=\sum_{j=1}^n \sum_{i=1}^m
\frac{\partial f_i}{\partial z_j}b_i\frac{\partial}{\partial
\eta_j}}$}$$ et de la graduation de Hodge, définie par
$deg(z_i)=0,\ deg(\eta_i)=1,\ deg(\alpha_j)=-1,\
deg(b_j)=2.$\\

\noindent On peut alors énoncer le théorème principal qui permet
le calcul de la cohomologie de Hochschild :

\begin{The}\label{kont}(Kontsevich)$\\$
Sous les hypothèses précédentes, la cohomologie de Hochschild de
$A$ est isomorphe à la cohomologie du complexe $(\widetilde{T},\
d_{\widetilde{T}})$ défini par l'algèbre différentielle graduée $\widetilde{T}$.\\
\end{The}

\noindent $\bullet$ Il n'y a pas d'élément de degré strictement négatif. On a donc le complexe suivant :\\

$$\xymatrix{\widetilde{T}(0) \ar@{->}[r]^{\widetilde{0}} &
\widetilde{T}(1) \ar@{->}[r]^{d_{\widetilde{T}}^{(1)}}
 & \widetilde{T}(2) \ar@{->}[r]^{d_{\widetilde{T}}^{(2)}}
 & \widetilde{T}(3) \ar@{->}[r]^{d_{\widetilde{T}}^{(3)}}
 & \widetilde{T}(4) \ar@{->}[r]^{d_{\widetilde{T}}^{(4)}}
 & \dots }$$\\

\noindent Pour chaque degré $p$, on choisit une base
$\mathcal{B}_p$ de $\widetilde{T}(p)$. Par exemple pour $p=0\dots 3$, on prend :\\

\noindent $\displaystyle{\widetilde{T}(0)=A}$\\
$\displaystyle{\widetilde{T}(1)=A\eta_1\oplus\dots\oplus A\eta_n}$\\
$\displaystyle{\widetilde{T}(2)=A b_1\oplus\dots\oplus A b_m\oplus \bigoplus_{i<j}A\,\eta_i \eta_j}$\\
$\displaystyle{\widetilde{T}(3)=\bigoplus_{\substack{i=1\dots m\\j=1\dots n}}A\, b_i \eta_j\oplus\bigoplus_{i<j<k}A\,\eta_i \eta_j \eta_k}$\\

\noindent On peut alors expliciter les matrices
$\displaystyle{Mat_{\mathcal{B}_p,\,\mathcal{B}_{p+1}}(d_{\widetilde{T}}^{(p)})}$.\\

\noindent $\bullet$ On note $p\ :\
\mathbb{C}[\mathbf{z}]\rightarrow A=\mathbb{C}[\mathbf{z}]/\langle
f_1,\dots,\ f_m\rangle$ la
projection canonique.\\
Pour tout idéal $J$ de $\mathbb{C}[\mathbf{z}]$, on notera $J_A$
l'image de cet idéal par la projection canonique.\\
De même, si $(g_1,\dots,\ g_r)\in A^r$ on notera $\langle
g_1,\dots,\ g_r\rangle _A$ l'idéal de $A$ engendré par
$(g_1,\dots,\ g_r)$.\\
Par ailleurs, si $g\in \mathbb{C}[\mathbf{z}]$, et si $J$ est un
idéal de $\mathbb{C}[\mathbf{z}]$, alors on note
$$Ann_J(g):=\{h\in \mathbb{C}[\mathbf{z}]\ /\ hg=0\mod J\}.$$
En particulier, $g$ ne divise pas $0$ dans
$\mathbb{C}[\mathbf{z}]/J$ si et seulement si $Ann_J(g)=J$.\\
Enfin, pour tout polynôme $g\in \mathbb{C}[\mathbf{z}]$, on note
$\nabla g$ son gradient.\\

\subsection{\textsf{Cas particulier où $n=1$ et $m=1$}}

\noindent $\bullet$ Dans le cas où $n=1$ et $m=1$, d'après ce qui
précède, on a pour $p\in \mathbb{N}^*$,
\begin{center}
\fbox{$\displaystyle{\widetilde{T}(2p)=A b_1^p}$} et
\fbox{$\displaystyle{\widetilde{T}(2p+1)=A b_1^p \eta_1.}$}
\end{center}
\noindent On en déduit

\noindent $H^0=A$, $H^1=\{g_1\eta_1\ /\ g_1\in A\ \textmd{et}\ g_1\,\partial_{z_1}f_1=0\}$\\
et $\forall\ p\in \mathbb{N}^*$, $H^{2p}=\displaystyle{\frac{Ab_1^p}{\{g_1 (\partial_{z_1}f_1) b_1^p\ /\ g_1\in A\}}}$, et $H^{2p+1}=\{g_1\,b_1^p\eta_1 \ /\ g_1\in A\ \textmd{et}\ g_1\,\partial_{z_1}f_1=0\}$.\\

\noindent $\bullet$ Si maintenant $f_1=z_1^k$, alors\\

\noindent $H^0=A=\mathbb{C}[z_1]\ /\ \langle z_1^k\rangle\simeq \mathbb{C}^{k-1}$\\
$H^1=\{g_1\eta_1\ /\ g_1\in A\ \textmd{et}\ kg_1z_1^{k-1}=0\}\simeq \mathbb{C}^{k-1}$\\
et $\forall\ p\in \mathbb{N}^*$,
$H^{2p}=\displaystyle{\frac{Ab_1^p}{\{g_1 (kz_1^{k-1}) b_1^p\ /\
g_1\in A\}}}\simeq \mathbb{C}^{k-1}$\\et
$H^{2p+1}=\{g_1\,b_1^p\eta_1 \ /\ g_1\in A\ \textmd{et}\ kg_1z_1^{k-1}=0\}\simeq \mathbb{C}^{k-1}$.\\

\section{\textsf{Cas $n=2,\ m=1$. --- courbes singulières du plan}}

\subsection{\textsf{Description des espaces de cohomologie}}

\noindent On utilise le théorème \ref{kont} pour calculer la
cohomologie de Hochschild de $A$. On commence par expliciter les
cochaînes et
les différentielles.\\

\noindent $\bullet$ Les différents espaces du complexe sont donnés
par

$$\begin{array}{l|l}
  \widetilde{T}(0)=A & \widetilde{T}(5)=A b_1^2\eta_1\oplus A b_1^2\eta_2 \\
  \widetilde{T}(1)=A\eta_1\oplus A\eta_2 & \widetilde{T}(6)=A b_1^3\oplus A b_1^2\eta_1\eta_2 \\
  \widetilde{T}(2)=A b_1\oplus A\eta_1\eta_2 & \widetilde{T}(7)=A b_1^3\eta_1\oplus A b_1^3\eta_2 \\
  \widetilde{T}(3)=A b_1\eta_1\oplus A b_1\eta_2 & \widetilde{T}(8)=A b_1^4\oplus A b_1^3\eta_1\eta_2 \\
  \widetilde{T}(4)=A b_1^2\oplus A b_1\eta_1\eta_2 & \widetilde{T}(9)=A b_1^4 \eta_1\oplus A b_1^4\eta_2, \\
\end{array}$$

\noindent ie, dans le cas général, pour $p\in \mathbb{N}^*$,
\fbox{$\widetilde{T}(2p)=A b_1^p\oplus A b_1^{p-1}\eta_1\eta_2$}
et \fbox{$\widetilde{T}(2p+1)=A b_1^p
\eta_1\oplus A b_1^p\eta_2$}.\\

\noindent On a
$\frac{\partial}{\partial\eta_k}(\eta_k\wedge\eta_l)=1\wedge\eta_l=-\eta_l\wedge1$,
donc $d_{\widetilde{T}}^{(2)}(\eta_k\eta_l)=-\frac{\partial
f_1}{\partial z_k} b_1\eta_l+\frac{\partial
f_1}{\partial z_l} b_1\eta_k$.\\
On note désormais $\frac{\partial}{\partial_{z_j}}=\partial_{z_j}=\partial_j$.\\

\noindent Les matrices de $d_{\widetilde{T}}$ sont donc données
par
\begin{center}
$\begin{array}{l}
 Mat_{\mathcal{B}_{2p},\mathcal{B}_{2p+1}}(d_{\widetilde{T}}^{(2p)})=\left(
\begin{array}{cc}
  0 & \partial_{2}f_1 \\
  0 & -\partial_{1}f_1 \\
\end{array}
\right) \\
  Mat_{\mathcal{B}_{2p+1},\mathcal{B}_{2p+2}}(d_{\widetilde{T}}^{(2p+1)})=\left(
\begin{array}{cc}
  \partial_{1}f_1 & \partial_{2}f_1 \\
  0 & 0 \\
\end{array}
\right). \\
\end{array}$
\end{center}

\noindent $\bullet$ On en déduit une expression plus simple des espaces de cohomologie :\\

\noindent $H^0=A$\\
$H^1=\{g_1\eta_1+g_2\eta_2\ /\ (g_1,\,g_2)\in A^2\ \textmd{et}\
g_1\,\partial_{1}f_1+g_2\,\partial_{2}f_1=0\}\simeq\left\{\mathbf{g}=\left(
\begin{array}{c}
  g_1 \\
  g_2 \\
\end{array}
\right)\in A^2\ /\ \mathbf{g}\cdot\nabla f_1=0\right\}$\\
$\forall\ p\in \mathbb{N}^*$,\\
$\begin{array}{rcl} H^{2p} & = & \frac{\{g_1 b_1^p+g_2
b_1^{p-1}\eta_1\eta_2\ /\ (g_1,\,g_2)\in A^2\ \textmd{et}\
g_2\,\partial_{1}f_1=g_2\,\partial_{2}f_1=0\}}{\{(g_1\,\partial_{1}f_1+g_2\,\partial_{2}f_1)b_1^p\
/\ (g_1,\,g_2)\in A^2\}}\simeq\frac{\left\{\mathbf{g}=\left(
\begin{array}{c}
  g_1 \\
  g_2 \\
\end{array}
\right)\in A^2\ /\
g_2\,\partial_{1}f_1=g_2\,\partial_{2}f_1=0\right\}}{\left\{\left(
\begin{array}{c}
  \mathbf{g}\cdot\nabla f_1 \\
  0 \\
\end{array}
\right)\ /\ \mathbf{g}\in A^2\right\}} \\
 & \simeq & \frac{A}{\langle \partial_{1}f_1,\ \partial_{2}f_1\rangle_A}\oplus \{g\in A\ /\ g\,\partial_{1}f_1=g\,\partial_{2}f_1=0\}\\
\end{array}$\\
$H^{2p+1}=\frac{\{g_1 b_1^p\eta_1+g_2 b_1^p\eta_2\ /\
(g_1,\,g_2)\in A^2\ \textmd{et}\
g_1\,\partial_{1}f_1+g_2\,\partial_{2}f_1=0\}}{\{g_2(\partial_{2}f_1b_1^p\eta_1-\partial_{1}f_1b_1^p\eta_2)\
/\ g_2\in A\}}\simeq\frac{\left\{\mathbf{g}=\left(
\begin{array}{c}
  g_1 \\
  g_2 \\
\end{array}
\right)\in A^2\ /\ \mathbf{g}\cdot\nabla
f_1=0\right\}}{\left\{g_2\left(
\begin{array}{c}
  \partial_{2}f_1 \\
  -\partial_{1}f_1 \\
\end{array}
\right)\ /\ g_2\in A\right\}}$.\\

\noindent Il reste à déterminer explicitement ces espaces. C'est l'objet des deux paragraphes suivants.\\

\subsection{\textsf{Calculs explicites dans le cas particulier où $f_1$ est à variables séparées}}

\noindent Dans ce paragraphe, on considère le polynôme
$f_1=a_1z_1^k+a_2z_2^l$, avec $2\leq l\leq k$ et
$(a_1,\,a_2)\in(\mathbb{C^*})^2$.

\noindent Les dérivées partielles de $f_1$ sont
$\partial_1f_1=ka_1z_1^{k-1}$ et
$\partial_2f_1=la_2z_2^{l-1}$.\\

\noindent $\bullet$ On a déjà $$H^0=\mathbb{C}[z_1,\,z_2]/\langle
a_1z_1^k+a_2z_2^l\rangle.$$\\

\noindent $\bullet$ De plus, comme $f_1$ est quasi-homogène, la
formule d'Euler donne
$\displaystyle{\frac{1}{k}x_1\partial_1f_1+\frac{1}{l}x_2\partial_2f_1=f_1}$.
Ainsi, on a l'inclusion $\langle f_1\rangle\subset\langle
\partial_1f_1,\,\partial_2f_1\rangle$, donc $\frac{A}
{\langle \partial_{z_1}f_1,\ \partial_{z_2}f_1\rangle_A}\simeq
\frac{\mathbb{C}[z_1,\,z_2]}{\langle \partial_{z_1}f_1,\
\partial_{z_2}f_1\rangle}\simeq Vect\left(z_1^iz_2^j\ /\ i\in\llbracket 0,\,k-2\rrbracket,\
j\in\llbracket 0,\,l-2\rrbracket\right)$.\\
Or $\partial_1f_1$ et $f_1$ sont premiers entre eux, de même que
$\partial_2f_1$ et $f_1$, donc si $g\in A$ vérifie
$g\partial_1f_1=0 \mod \langle f_1\rangle$, alors $g\in \langle
f_1\rangle$, ie $g$ est nul dans $A$.\\
Ainsi, $$H^{2p}\simeq Vect\left(z_1^iz_2^j\ /\ i\in\llbracket
0,\,k-2\rrbracket,\
j\in\llbracket 0,\,l-2\rrbracket\right)\simeq \mathbb{C}^{(k-1)(l-1)}.$$\\

\noindent $\bullet$ On détermine maintenant l'ensemble
$\left\{\mathbf{g}=\left(
\begin{array}{c}
  g_1 \\
  g_2 \\
\end{array}
\right)\in A^2\ /\ \mathbf{g}\cdot\nabla f_1=0\right\}$ :\\
On a d'abord $\langle f_1,\,\partial_1f_1\rangle=\langle
a_1z_1^k+a_2z_2^l,\,z_1^{k-1}\rangle=\langle
z_2^l,\,z_1^{k-1}\rangle$. Les seuls monômes qui ne sont pas dans
cet idéal sont donc les éléments $z_1^iz_2^j$ avec $i\in\llbracket
0,\,k-2\rrbracket$ et $j\in\llbracket 0,\,l-1\rrbracket$.\\
Tout polynôme $P\in \mathbb{C}[\mathbf{z}]$ peut donc s'écrire
sous la forme
$$\displaystyle{P=\alpha f_1+\beta
\partial_1f_1+\sum_{\substack{i=0\dots k-2 \\ j=0\dots
l-1}}a_{ij}z_1^iz_2^j}.$$\\
Les polynômes $P\in \mathbb{C}[\mathbf{z}]$ tels que
$P\partial_2f_1\in \langle f_1,\,\partial_1f_1\rangle$ sont donc
les éléments $$\displaystyle{P=\alpha f_1+\beta
\partial_1f_1+\sum_{\substack{i=0\dots k-2 \\ j=1\dots
l-1}}a_{ij}z_1^iz_2^j}.$$ On a ainsi calculé $Ann_{\langle f_1,\,\partial_1f_1\rangle}(\partial_2f_1)$.\\
L'équation
\begin{equation}\label{21eq1}
\mathbf{g}\cdot\nabla f_1=0 \mod \langle f_1\rangle
\end{equation}
entraîne
\begin{equation}\label{21eq2}
g_2\partial_2f_1=0 \mod \langle f_1,\,\partial_1f_1\rangle,
\end{equation}
ie $g_2\in Ann_{\langle
f_1,\,\partial_1f_1\rangle}(\partial_2f_1)$, ie encore
\begin{equation}\label{21eq3}
g_2=\alpha f_1+\beta\partial_1f_1+\sum_{\substack{i=0\dots k-2
\\ j=1\dots
l-1}}a_{ij}z_1^iz_2^j,
\end{equation}
avec $(\alpha,\,\beta)\in \mathbb{C}[\mathbf{z}]^2$.\\
Il s'ensuit que
\begin{equation}\label{21eq4}
g_1\partial_1f_1+\alpha
f_1\partial_2f_1+\beta\partial_1f_1\partial_2f_1+\sum_{\substack{i=0\dots
k-2 \\ j=1\dots l-1}}a_{ij}z_1^iz_2^j \partial_2f_1\in \langle
f_1\rangle.
\end{equation}
Et, avec l'égalité
$z_2\partial_2f_1=lf_1-\frac{l}{k}z_1\partial_1f_1$,
\begin{equation}\label{21eq5}
\partial_1f_1\left(g_1+\beta\partial_2f_1-\frac{l}{k}\sum_{\substack{i=0\dots
k-2 \\ j=1\dots l-1}}a_{ij}z_1^{i+1}z_2^{j-1}\right)\in \langle
f_1\rangle.
\end{equation}
Comme $f_1$ et $\partial_1f_1$ sont premiers entre eux, cette
équation équivaut à
$$g_1=-\beta\partial_2f_1+\frac{l}{k}\sum_{\substack{i=0\dots k-2 \\
j=1\dots l-1}}a_{ij}z_1^{i+1}z_2^{j-1}+\delta f_1,$$ avec
$\delta\in \mathbb{C}[\mathbf{z}]$.\\
On vérifie ensuite que les éléments $g_1$ et $g_2$ ainsi obtenus
sont bien solutions de l'équation (\ref{21eq1}).\\
Finalement, on a
\begin{small}
$$\left\{\mathbf{g}\in A^2\ /\
\mathbf{g}\cdot\nabla f_1=0 \right\}=\left\{ \left(%
\begin{array}{c}
  \alpha \\
  \delta \\
\end{array}%
\right)f_1-\beta\left(%
\begin{array}{c}
  \partial_2f_1 \\
  -\partial_1f_1 \\
\end{array}%
\right)+\sum_{\substack{i=0\dots k-2 \\ j=1\dots
l-1}}a_{ij}z_1^{i}z_2^{j-1}\left(%
\begin{array}{c}
  \frac{l}{k}z_1 \\
  z_2 \\
\end{array}%
\right)\ \Big/\ (\alpha,\,\beta,\,\delta)\in
\mathbb{C}[\mathbf{z}]^3\ \textrm{et}\ a_{ij}\in
\mathbb{C}\right\}.$$
\end{small}
On en déduit aussitôt les espaces de cohomologie d'indices impairs
:
$$\begin{array}{rcl}
\forall\ p\geq 1,\ H^{2p+1} & \simeq & \mathbb{C}^{(k-1)(l-1)} \\
 H^{1} & \simeq & \mathbb{C}^{(k-1)(l-1)}\oplus \mathbb{C}[z_1,\,z_2]/\langle a_1z_1^k+a_2z_2^l \rangle.\\
\end{array}$$

\begin{Rq}$\\$
On obtient en particulier la cohomologie pour les cas où
$f_1=z_1^{k+1}+z_2^2$, $f_1=z_1^3+z_2^4$ et $f_1=z_1^3+z_2^5$, cas
qui correspondent respectivement aux fonctions quasi-homogènes de
types $A_k$, $E_6$ et $E_8$ données dans \cite{AVGZ86} p. 181.\\
\end{Rq}

\noindent On regroupe ces trois cas particuliers dans le tableau
ci-dessous
:\\

\begin{tabular}{|l||l|l|l|l|}
  \hline
    & $H^0$ & $H^1$ & $H^{2p}$ & $H^{2p+1}$ \\
  \hline\hline
  $A_k$ & $\mathbb{C}[\mathbf{z}]\ /\ \langle z_1^{k+1}+z_2^2\rangle$ & $\mathbb{C}[\mathbf{z}]\ /\ \langle z_1^{k+1}+z_2^2\rangle\oplus \mathbb{C}^k$ & $\mathbb{C}^k$ & $\mathbb{C}^k$
  \\ \hline
  $E_6$ & $\mathbb{C}[\mathbf{z}]\ /\ \langle z_1^3+z_2^4\rangle$ & $\mathbb{C}[\mathbf{z}]\ /\ \langle z_1^3+z_2^4\rangle\oplus \mathbb{C}^6$ & $\mathbb{C}^6$ & $\mathbb{C}^6$
  \\ \hline
  $E_8$ & $\mathbb{C}[\mathbf{z}]\ /\ \langle z_1^3+z_2^5\rangle$ & $\mathbb{C}[\mathbf{z}]\ /\ \langle z_1^3+z_2^5\rangle\oplus \mathbb{C}^8$ & $\mathbb{C}^8$ & $\mathbb{C}^8$ \\
  \hline
\end{tabular}\\

\noindent \\ Les cas où $f_1=z_1^2z_2+z_2^{k-1}$ et
$f_1=z_1^3+z_1z_2^3$, ie respectivement $D_k$ et $E_7$, sont
étudiés dans le paragraphe suivant.\\

\subsection{\textsf{Calculs explicites pour $D_k$ et $E_7$}}

\noindent Pour étudier ces cas particuliers, on utilise le
résultat suivant sur les bases de Gröbner :

\begin{Def}$\\$
Pour $g\in \mathbb{C}[\mathbf{z}]$, on note $lt(g)$ son terme
dominant (pour l'ordre lexicographique).\\
Soit $J$ un idéal de $\mathbb{C}[\mathbf{z}]$ et
$G_J:=[g_1,\,\dots,\,g_r]$ une base de Gröbner de $J$. Un polynôme
$p\in \mathbb{C}[\mathbf{z}]$ est réduit relativement à $G_J$ s'il
est nul ou bien si aucun terme de $p$ n'est divisible par le terme
dominant $lt(g_j)$ de l'un des éléments de $G_J$.\\
L'ensemble des termes $G_J-$standards est l'ensemble des monômes
de $\mathbb{C}[\mathbf{z}]$ privé de l'ensemble des termes
dominants $lt(f)$ des polynômes $f\in J\backslash\{0\}$.\\
\end{Def}

\begin{The}\label{Macaulay}(Macaulay)$\\$
L'ensemble des termes $G_J-$standards forme une base de l'espace
vectoriel quotient $\mathbb{C}[\mathbf{z}]\ /\ J$.\\
\end{The}

\subsubsection{\textsf{Cas de $f_1=z_1^2z_2+z_2^{k-1}$, ie $D_k$}}

\noindent On a ici $f_1=z_1^2z_2+z_2^{k-1}$,
$\partial_1f_1=2z_1z_2$ et
$\partial_2f_1=z_1^2+(k-1)z_2^{k-2}$.\\
Une base de Gröbner de l'idéal $\langle
f_1,\,\partial_2f_1\rangle$ est
$B:=[b_1,\,b_2]=[z_1^2+(k-1)z_2^{k-2},\,z_2^{k-1}]$.\\
L'ensemble des termes standards est donc $\{z_1^iz_2^j\ /\ i\in
\{0,\,1\}\ \textrm{et}\ j\in \llbracket 0,\,k-2\rrbracket\}$.\\
On peut alors résoudre l'équation $p\,\partial_1f_1=0$ dans
$\mathbb{C}[\mathbf{z}]\ /\ \langle f_1,\,\partial_2f_1\rangle$.\\
En effet, en écrivant $p:=\displaystyle{\sum_{\substack{i=0,1} \\
j=0\dots k-2}a_{ij}z_1^iz_2^j}$, l'équation devient $$\displaystyle{q:=\sum_{\substack{i=0,1} \\
j=0\dots k-2}a_{ij}z_1^{i+1}z_2^{j+1}\in \langle
f_1,\,\partial_2f_1\rangle}.$$ On cherche donc la forme normale de
l'élément $q$ modulo l'idéal $\langle
f_1,\,\partial_2f_1\rangle$.\\
La division multivariée de $q$ par $B$ s'écrit $q=q_1b_1+q_2b_2+r$
avec $r=\sum_{j=0}^{k-3}a_{0,j}z_1z_2^{j+1}$.\\
La solution est donc
$$p=a_{0,k-2}z_2^{k-2}+\sum_{j=0}^{k-2}a_{1,j}z_1z_2^j.$$\\
Or l'équation
\begin{equation}\label{21beq1}
\mathbf{g}\cdot\nabla f_1=0 \mod \langle f_1\rangle
\end{equation}
entraîne
\begin{equation}\label{21beq2}
g_1\partial_1f_1=0 \mod \langle f_1,\,\partial_2f_1\rangle,
\end{equation}
ie
\begin{equation}\label{21beq3}
g_1=\alpha
f_1+\beta\partial_2f_1+az_2^{k-2}+\sum_{j=0}^{k-2}b_jz_1z_2^j,
\end{equation}
avec $(\alpha,\,\beta)\in \mathbb{C}[\mathbf{z}]^2$ et $a,\,b_j\in \mathbb{C}$.\\
D'où
\begin{equation}\label{21beq4}
g_2\partial_2f_1+\beta\partial_1f_1\partial_2f_1+az_2^{k-2}\partial_1f_1+
\sum_{j=0}^{k-2}b_jz_1z_2^j\partial_1f_1\in \langle f_1\rangle.
\end{equation}
Et, avec les égalités
$z_2^{k-1}=\frac{1}{2-k}(f_1-z_2\partial_2f_1)=-\frac{1}{2-k}z_2\partial_2f_1\mod
\langle f_1\rangle$, et
$\frac{k-2}{2}z_1\partial_1f_1+z_2\partial_2f_1=(k-1)f_1$ (Euler),
on obtient
\begin{equation}\label{21beq5}
\partial_2f_1\left(g_2+\beta\partial_1f_1 -\frac{2a}{2-k}z_1z_2+
\sum_{j=0}^{k-2}b_j\frac{2}{2-k}z_2^{j+1}\right)\in \langle
f_1\rangle.
\end{equation}
Mais $f_1$ et $\partial_2f_1$ sont premiers entre eux, donc
$$g_2=-\beta\partial_1f_1+\frac{2a}{2-k}z_1z_2-
\sum_{j=0}^{k-2}b_j\frac{2}{2-k}z_2^{j+1}+\delta f_1,$$ avec
$\delta\in \mathbb{C}[\mathbf{z}]$.\\
Donc
\begin{scriptsize}
$$\left\{\mathbf{g}\in A^2\ /\
\mathbf{g}\cdot\nabla f_1=0 \right\}=\left\{ \left(%
\begin{array}{c}
  \alpha \\
  \delta \\
\end{array}%
\right)f_1+\beta\left(%
\begin{array}{c}
  \partial_2f_1 \\
  -\partial_1f_1 \\
\end{array}%
\right)+\left(%
\begin{array}{c}
  z_2^{k-2} \\
  \frac{2}{2-k}z_1z_2 \\
\end{array}%
\right)+\sum_{j=0}^{k-2}b_jz_2^j\left(%
\begin{array}{c}
  z_1 \\
  -\frac{2}{2-k}z_2 \\
\end{array}%
\right)\ \Big/\ (\alpha,\,\beta,\,\delta)\in
\mathbb{C}[\mathbf{z}]^3\ \textrm{et}\ a,\,b_j\in
\mathbb{C}\right\}.$$
\end{scriptsize}\\

\noindent Par ailleurs, une base de Gröbner de $\langle
\partial_1f_1,\,\partial_2f_1\rangle$ est
$[z_1^2+(k-1)z_2^{k-2},\,z_1z_2,\,z_2^{k-1}]$,\\
donc
$\mathbb{C}[\mathbf{z}]\ /\ \langle
\partial_1f_1,\,\partial_2f_1\rangle\simeq Vect\left(z_1,\,1,\,z_2\,\dots,\,z_2^{k-2}\right)$.\\

\noindent En résumé, \\

\noindent \begin{tabular}{|l|}
  \hline
  $H^0=\mathbb{C}[\mathbf{z}]\ /\ \langle
z_1^2z_2+z_2^{k-1}\rangle$\\
$H^1\simeq \mathbb{C}[\mathbf{z}]\ /\ \langle
z_1^2z_2+z_2^{k-1}\rangle\oplus \mathbb{C}^k$\\
$H^{2p}\simeq \mathbb{C}^k$\\
$H^{2p+1}\simeq \mathbb{C}^k$. \\
  \hline
\end{tabular}\\

\subsubsection{\textsf{Cas de $f_1=z_1^3+z_1z_2^3$, ie $E_7$}}

\noindent On a ici $\partial_1f_1=3z_1^2+z_2^3$ et
$\partial_2f_1=3z_1z_2^2$.\\
Une base de Gröbner de l'idéal $\langle
f_1,\,\partial_1f_1\rangle$ est
$[3z_1^2+z_2^3,\,z_1z_2^3,\,z_2^6]$, et une base de Gröbner de
$\langle
\partial_1f_1,\,\partial_2f_1\rangle$ est
$[3z_1^2+z_2^3,\,z_1z_2^2,\,z_2^5]$.
\\Par une démonstration analogue, on obtient :\\

\noindent\begin{tabular}{|l|}
  \hline
$H^0=\mathbb{C}[\mathbf{z}]\ /\ \langle
z_1^3+z_1z_2^3\rangle$\\
$H^1\simeq \mathbb{C}[\mathbf{z}]\ /\ \langle
z_1^3+z_1z_2^3\rangle\oplus \mathbb{C}^7$\\
$H^{2p}\simeq \mathbb{C}^7$\\
$H^{2p+1}\simeq \mathbb{C}^7$. \\
  \hline
\end{tabular}\\

\section{\textsf{Cas $n=3,\ m=1$. --- surfaces de Klein}}

\subsection{\textsf{Surfaces de Klein}}

\noindent Etant donné un groupe fini $G$ agissant sur
$\mathbb{C}^n$, on lui fait correspondre, selon le programme
d'Erlangen de Klein, la variété quotient $\mathbb{C}^n/G$ : c'est
la variété dont les points sont les orbites sous l'action de $G$.
Les fonctions polynomiales sur cette variété sont les fonctions
polynomiales sur $\mathbb{C}^n$ invariantes par $G$.\\
Dans le cas de $\mathbf{SL}_2\mathbb{C}$, la théorie des
invariants permet d'associer à tout sous-groupe fini un polynôme.
Ainsi, à tout sous-groupe fini de $\mathbf{SL}_2\mathbb{C}$ est
associée la variété algébrique constituée des zéros de ce
polynôme, appelée surface de Klein.
\\On rappelle dans ce paragraphe
quelques résultats sur ces surfaces. Voir les références \cite{S77} et \cite{CCK99} pour plus de détails.\\

\begin{Pro}$\\$
Tout sous-groupe fini de $\mathbf{SL}_2\mathbb{C}$ est conjugué à l'un des groupes suivants :\\
$\bullet$ $A_n$ (cyclique), $n\geq 1$ ($|A_n|=n$)\\
$\bullet$ $D_n$ (diédral), $n\geq 1$ ($|D_n|=4n$)\\
$\bullet$ $E_6$ (tétraédral) ($|E_6|=24$)\\
$\bullet$ $E_7$ (octaédral) ($|E_7|=48$)\\
$\bullet$ $E_8$ (icosaédral) ($|E_8|=120$).\\
\end{Pro}

\begin{Pro}$\\$
Soit $G$ l'un des groupes de la liste précédente. L'anneau des
invariants est
$$\mathbb{C}[x,\,y]^G=\mathbb{C}[e_1,\,e_2,\,e_3]=\mathbb{C}[e_1,\,e_2]
\oplus e_3\mathbb{C}[e_1,\,e_2]\simeq
\mathbb{C}[z_1,\,z_2,\,z_3]/\langle f_1\rangle,$$ où les
invariants $e_j$ sont des polynômes homogènes, avec $e_1$ et $e_2$
algébriquement indépendants, et où $f_1$
est un polynôme quasi-homogène à singularité isolée en l'origine.\\
Ces polynômes sont donnés dans le tableau suivant.\\
\end{Pro}

\noindent\begin{tabular}{|l||l|l|l|} \hline $G$ & $e_1,\ e_2,\
e_3$ & $f_1$ & $\mathbb{C}[z_1,\,z_2,\,z_3]/\langle\partial_1
f_1,\
\partial_2 f_1,\ \partial_3 f_1\rangle$  \\ \hline
\hline $A_n$ &
\begin{minipage}{5cm}
\vspace{.2cm}$e_1=x^n$\\ $e_2=y^n$ \\$e_3=xy$ \vspace{.2cm}
\end{minipage} & $-n(z_1z_2-z_3^n)$ &
\begin{minipage}{4.5cm}
\vspace{.2cm}$Vect(1,\,z_1,\,\dots,\,z_3^{n-2})$\\
$\dim=n-1$ \vspace{.2cm}
\end{minipage}\\
\hline $D_n$ & \begin{minipage}{5cm} \vspace{.2cm}$e_1=x^2y^2$\\
$e_2=x^{2n}+(-1)^ny^{2n}$
\\$e_3=x^{2n+1}y+(-1)^{n+1}xy^{2n+1}$ \vspace{.2cm}
\end{minipage} &
\begin{minipage}{5cm}
\vspace{.2cm} \begin{small}
$\lambda_n(4z_1^{n+1}+(-1)^{n+1}z_1z_2^2+(-1)^nz_3^2)$ \end{small}
\\ avec $\lambda_n=2n(-1)^{n+1}$ \vspace{.2cm}
\end{minipage} &
\begin{minipage}{4.5cm}
\vspace{.2cm}$Vect(1,\,z_2,\,z_1,\,\dots,\,z_1^{n-1})$
\\ $\dim=n+1$ \vspace{.2cm}
\end{minipage}\\
\hline $E_6$ &
\begin{minipage}{5cm} \vspace{.2cm}$e_1=x^5y-xy^5$ \\
$e_2=14y^4x^4+x^8+y^8$\\
$e_3=33y^8x^4-y^{12}+33y^4x^8-x^{12}$ \vspace{.1cm}
\end{minipage} & $4(z_3^2-z_2^3+108z_1^4)$ &
\begin{minipage}{4.5cm}
\vspace{.2cm}$Vect(1,\,z_2,\,z_1,\,z_1z_2,\,z_1^2,\,z_1^2z_2)$ \\
$\dim=6$ \vspace{.2cm}
\end{minipage}\\
\hline $E_7$ &
\begin{minipage}{5.65cm} \vspace{.2cm}$e_1=14y^4x^4+x^8+y^8$\\
$e_2=-3y^{10}x^2+6y^6x^6-3y^2x^{10}$\\
$e_3=-34x^5y^{13}-yx^{17}+34y^5x^{13}+xy^{17}$ \vspace{.2cm}
\end{minipage} & $8(3z_3^2-12z_2^3+z_2z_1^3)$ &
\begin{minipage}{4.5cm}
\vspace{.2cm}$Vect(1,\,z_2,\,z_2^2,\,z_1,\,z_1z_2,\,z_1z_2^2,\,z_1^2)$ \\
$\dim=7$ \vspace{.2cm}
\end{minipage}\\
\hline $E_8$ & \begin{minipage}{5.8cm}
\vspace{.2cm}$\begin{array}{rl}
 e_1= & x^{11}y+11x^6y^6-xy^{11} \\
 e_2= & x^{20}-228x^{15}y^5+494x^{10}y^{10}\\
  & +228x^5y^{15}+y^{20} \\
 e_3= & x^{30}+522x^{25}y^5\\
  & -10\,005x^{20}y^{10}-10\,005x^{10}y^{20}\\
   & -522x^5y^{25}+y^{30} \\
  \end{array}$\vspace{.2cm}
\end{minipage}
 & $10(1\,728z_1^5+z_2^3-z_3^2)$ &
\begin{minipage}{4.5cm}
\vspace{.2cm}$Vect(z_1^iz_2^j)_{\substack{i=0\dots 3,\\ j=0\dots 1}}$ \\
$\dim=8$ \vspace{.2cm}
\end{minipage} \\
\hline
\end{tabular}\\

\noindent\\ On appelle surface de Klein l'hypersurface algébrique
définie par
$\{\mathbf{z}\in \mathbb{C}^3\ /\ f_1(\mathbf{z})=0\}.$\\

\begin{The}\label{Pich}(Pichereau)$\\$
On considère le crochet de Poisson défini sur
$\mathbb{C}[z_1,\,z_1,\,z_3]$ par $$\{\cdot\}_{f_1}=\partial_3
f_1\,
\partial_1\wedge\partial_2+\partial_1 f_1\,
\partial_2\wedge\partial_3+\partial_2 f_1\,
\partial_3\wedge\partial_1=i_{df_1}(\partial_1\wedge\partial_2\wedge\partial_3),$$ et on note $HP^*_{f_1}$ (resp.
$HP_*^{f_1}$) la cohomologie (resp. l'homologie) de Poisson pour
ce crochet. Sous les hypothèses précédentes, la cohomologie de
Poisson $HP^*_{f_1}$ et l'homologie de Poisson $HP_*^{f_1}$ de
$(\mathbb{C}[z_1,\,z_1,\,z_3]/\langle f_1\rangle,\
\{\cdot\}_{f_1})$ est donnée par
\begin{center}
$\begin{array}{l}
 HP^0_{f_1}=\mathbb{C},\ \  HP^1_{f_1}\simeq HP^2_{f_1}=\{0\} \\
 HP_0^{f_1}\simeq HP_2^{f_1}\simeq
\mathbb{C}[z_1,\,z_2,\,z_3]/\langle\partial_1 f_1,\
\partial_2 f_1,\ \partial_3 f_1\rangle \\
  \dim(HP_1^{f_1})=\dim(HP_0^{f_1})-1 \\
HP_j^{f_1}=HP_{f_1}^j=\{0\}\ \textrm{si}\ j\geq 3.\\
\end{array}$
\end{center}
\end{The}

\noindent L'algèbre $\mathbb{C}[x,\,y]$ est une algèbre de Poisson
pour le crochet symplectique standard que l'on note
$\{\cdot\}_{std}$. Comme $G$ est un sous-groupe du groupe
symplectique $\mathbf{Sp}_2\mathbb{C}$, l'algèbre des invariants
$\mathbb{C}[x,\,y]^G$ est une sous-algèbre de Poisson de
$\mathbb{C}[x,\,y]$. La propriété suivante permet alors de déduire
du théorème \ref{Pich} la cohomologie de
 Poisson et l'homologie de Poisson de $\mathbb{C}[x,\,y]^G$ pour le
 crochet symplectique standard.

\begin{Pro}$\\$
L'isomorphisme d'algèbres associatives $$\begin{array}{rcl}
  \pi:(\mathbb{C}[x,\,y]^G,\
\{\cdot\}_{std}) & \rightarrow &
(\mathbb{C}[z_1,\,z_1,\,z_3]/\langle f_1\rangle,\
\{\cdot\}_{f_1}) \\
  e_j & \mapsto & \overline{z_j} \\
\end{array}$$
 est un isomorphisme de Poisson.\\
\end{Pro}

\noindent Dans la suite, on va calculer la cohomologie de
Hochschild de $\mathbb{C}[z_1,\,z_1,\,z_3]/\langle f_1\rangle$. On
en déduira alors immédiatement la cohomologie de Hochschild de
$\mathbb{C}[x,\,y]^G$, grâce à l'isomorphisme d'algèbres associatives $\pi$.\\

\subsection{\textsf{Description des espaces de cohomologie}}

\noindent $\bullet$ Dans ce cas, on change l'ordre des vecteurs de
base : on prend $(\eta_1\eta_2,\ \eta_2\eta_3,\ \eta_3\eta_1)$ au
lieu de $(\eta_1\eta_2,\ \eta_1\eta_3,\ \eta_2\eta_3)$. Les
différents espaces du complexe sont alors donnés par

$$\begin{array}{l}
  \widetilde{T}(0)=A\\
  \widetilde{T}(1)=A\eta_1\oplus A\eta_2\oplus A\eta_3 \\
  \widetilde{T}(2)=A b_1\oplus A\eta_1\eta_2\oplus A\eta_2\eta_3\oplus A\eta_3\eta_1 \\
  \widetilde{T}(3)=A b_1\eta_1\oplus A b_1\eta_2\oplus A b_1\eta_3\oplus A \eta_1\eta_2\eta_3 \\
  \widetilde{T}(4)=A b_1^2\oplus A b_1\eta_1\eta_2\oplus A b_1\eta_2\eta_3\oplus A b_1\eta_3\eta_1  \\
\widetilde{T}(5)=A b_1^2\eta_1\oplus A b_1^2\eta_2\oplus A b_1^2\eta_3\oplus A b_1\eta_1\eta_2\eta_3\\
\end{array}$$

\noindent ie, dans le cas général, pour $p\in \mathbb{N}^*$,
 \fbox{$\widetilde{T}(2p)=A
b_1^p\oplus A b_1^{p-1}\eta_1\eta_2\oplus A
b_1^{p-1}\eta_2\eta_3\oplus A b_1^{p-1}\eta_3\eta_1$}\\
et
\fbox{$\widetilde{T}(2p+1)=A b_1^p
\eta_1\oplus A b_1^p\eta_2\oplus A b_1^p\eta_3\oplus A b_1^{p-1}\eta_1\eta_2\eta_3$}.\\

\noindent On a
$\frac{\partial}{\partial\eta_1}(\eta_1\wedge\eta_2\wedge\eta_3)=1\wedge\eta_2\wedge\eta_3=\eta_2\wedge\eta_3\wedge
1$, donc $d_{\widetilde{T}}^{(3)}(\eta_1\eta_2\eta_3)=
\frac{\partial f_1}{\partial z_1} b_1\eta_2\eta_3+\frac{\partial
f_1}{\partial z_2} b_1\eta_3\eta_1
+\frac{\partial f_1}{\partial z_3} b_1\eta_1\eta_2.$\\

\noindent Les matrices de $d_{\widetilde{T}}$ sont donc données
par
\begin{center}
$\begin{array}{c}
Mat_{\mathcal{B}_{1},\mathcal{B}_{2}}(d_{\widetilde{T}}^{(1)})=\left(
\begin{array}{ccc}
  \partial_{z_1}f_1 & \partial_{z_2}f_1 & \partial_{z_3}f_1 \\
  0 & 0 & 0 \\
    0 & 0 & 0 \\
      0 & 0 & 0 \\
\end{array}
\right) \\
\forall\ p\in \mathbb{N}^*,\
Mat_{\mathcal{B}_{2p},\mathcal{B}_{2p+1}}(d_{\widetilde{T}}^{(2p)})=\left(
\begin{array}{cccc}
0 & \partial_{z_2}f_1 & 0 & -\partial_{z_3}f_1 \\
  0 & -\partial_{z_1}f_1 & \partial_{z_3}f_1 & 0 \\
    0 & 0 & -\partial_{z_2}f_1 & \partial_{z_1}f_1 \\
      0 & 0 & 0  & 0\\
\end{array}
\right)
\\ \forall\ p\in \mathbb{N}^*,\
Mat_{\mathcal{B}_{2p+1},\mathcal{B}_{2p+2}}(d_{\widetilde{T}}^{(2p+1)})=\left(
\begin{array}{cccc}
\partial_{z_1}f_1 & \partial_{z_2}f_1 & \partial_{z_3}f_1 & 0 \\
  0 & 0 & 0 & \partial_{z_3}f_1\\
    0 & 0 & 0  & \partial_{z_1}f_1\\
      0 & 0 & 0  & \partial_{z_2}f_1\\
\end{array}
\right). \\
\end{array}$
\end{center}

\noindent $\bullet$ On en déduit\\

\noindent $H^0=A$\\
$H^1=\{g_1\eta_1+g_2\eta_2+g_3\eta_3\ /\ (g_1,\,g_2,\,g_3)\in A^3\
\textmd{et}\
g_1\,\partial_{z_1}f_1+g_2\,\partial_{z_2}f_1+g_3\,\partial_{z_3}f_1=0\}\simeq\left\{\mathbf{g}=\left(
\begin{array}{c}
  g_1 \\
  g_2 \\
  g_3 \\
\end{array}
\right)\in A^3\ /\ \mathbf{g}\cdot\nabla f_1=0\right\}$\\

$\begin{array}{rcl}
  H^{2} & = & \frac{\{g_0 b_1+g_3\eta_1\eta_2+g_1\eta_2\eta_3+g_2\eta_3\eta_1\
/\ (g_0,\,g_1,\,g_2\,g_3)\in A^4\ \textmd{et}\
g_3\,\partial_{z_2}f_1-g_2\,\partial_{z_3}f_1=g_1\,\partial_{z_3}f_1-g_3\,\partial_{z_1}f_1=g_2\,\partial_{z_1}f_1-g_1\,\partial_{z_2}f_1=0\}}
{\{(g_1\,\partial_{z_1}f_1+g_2\,\partial_{z_2}f_1+g_3\,\partial_{z_3}f_1)b_1\
/\ (g_1,\,g_2,\,g_3)\in A^3\}} \\
    & \simeq & \frac{\left\{\mathbf{g}=\left(
\begin{array}{c}
 g_0\\
  g_1 \\
  g_2 \\
 g_3 \\
\end{array}
\right)\in A^4\ \Big/\ \nabla f_1\wedge \left(
\begin{array}{c}
  g_1 \\
  g_2 \\
 g_3 \\
\end{array}
\right)=0\right\}}{\left\{\left(
\begin{array}{c}
  \mathbf{g}\cdot\nabla f_1 \\
  \mathbf{0}_{3,1} \\
\end{array}
\right)\ /\ \mathbf{g}\in A^3\right\}} \\
    & \simeq & \frac{A}{\langle \partial_{z_1}f_1,\ \partial_{z_2}f_1,\ \partial_{z_3}f_1\rangle_A}\oplus \{\mathbf{g}\in A^3\ /\ \nabla f_1\wedge \mathbf{g}=0\}\\
\end{array}$\\

\noindent $\forall\ p\geq 2,$\\
$\begin{array}{rcl}
  H^{2p} & = & \frac{\left\{g_0 b_1^p+g_3b_1^{p-1}\eta_1\eta_2+g_1b_1^{p-1}\eta_2\eta_3+g_2b_1^{p-1}\eta_3\eta_1\
\Big/\ \substack{(g_0,\,g_1,\,g_2\,g_3)\in A^4\ \textmd{et}\ \\
g_3\,\partial_{z_2}f_1-g_2\,\partial_{z_3}f_1=g_1\,\partial_{z_3}f_1-g_3\,\partial_{z_1}f_1=g_2\,\partial_{z_1}f_1-g_1\,\partial_{z_2}f_1=0}\right\}}
{\{(g_1\,\partial_{z_1}f_1+g_2\,\partial_{z_2}f_1+g_3\,\partial_{z_3}f_1)b_1^p+g_0(\partial_{z_3}f_1\,b_1^{p-1}\eta_1\eta_2
+\partial_{z_1}f_1\,b_1^{p-1}\eta_2\eta_3+\partial_{z_2}f_1\,b_1^{p-1}\eta_3\eta_1)\
/\ (g_0,\,g_1,\,g_2,\,g_3)\in A^3\}} \\
    & \simeq & \frac{\left\{\mathbf{g}=\left(
\begin{array}{c}
 g_0\\
  g_1 \\
  g_2 \\
 g_3 \\
\end{array}
\right)\in A^4\ \Big/\ \nabla f_1\wedge \left(
\begin{array}{c}
  g_1 \\
  g_2 \\
 g_3 \\
\end{array}
\right)=0\right\}}{\left\{\left(
\begin{array}{c}
  \mathbf{g}\cdot\nabla f_1 \\
  g_0\,\partial_{z_1}f_1 \\
  g_0\,\partial_{z_2}f_1 \\
  g_0\,\partial_{z_3}f_1 \\
\end{array}
\right)\ /\ \mathbf{g}\in A^3\ \textmd{et}\ g_0\in A\right\}} \\
    & \simeq & \frac{A}{\langle \partial_{z_1}f_1,\ \partial_{z_2}f_1,\ \partial_{z_3}f_1\rangle_A}\oplus
     \frac{\{\mathbf{g}\in A^3\ /\ \nabla f_1\wedge \mathbf{g}=0\}}{\left\{g\nabla f_1\ /\ g\in A\right\}}\\
\end{array}$\\

\noindent $\forall\ p\in \mathbb{N}^*,$\\
$\begin{array}{rcl}
  H^{2p+1} & = & \frac{\left\{g_1 b_1^p\eta_1+g_2 b_1^p\eta_2+g_3 b_1^p\eta_3+g_0b_1^{p-1}\eta_1\eta_2\eta_3\
\Big/\ \substack{(g_0,\,g_1,\,g_2\,g_3)\in A^4\ \textmd{et}\
g_1\,\partial_{z_1}f_1+g_2\,\partial_{z_2}f_1+g_3\,\partial_{z_3}f_1=0
\\ g_0\,\partial_{z_3}f_1=g_0\,\partial_{z_1}f_1=g_0\,\partial_{z_2}f_1=0}\right\}}
{\{(g_3\,\partial_{z_2}f_1-g_2\,\partial_{z_3}f_1)b_1^p\eta_1+(g_1\,\partial_{z_3}f_1-g_3\,\partial_{z_1}f_1)b_1^p\eta_2
+(g_2\,\partial_{z_1}f_1-g_1\,\partial_{z_2}f_1)b_1^p\eta_3\
/\ (g_1,\,g_2,\,g_3)\in A^3\}} \\
    & \simeq & \frac{\left\{\mathbf{g}=\left(
\begin{array}{c}
 g_1\\
  g_2 \\
  g_3 \\
 g_0 \\
\end{array}
\right)\in A^4\ \Big/\ \nabla f_1\cdot \left(
\begin{array}{c}
  g_1 \\
  g_2 \\
 g_3 \\
\end{array}
\right)=0\ \textmd{et}\
g_0\,\partial_{z_3}f_1=g_0\,\partial_{z_1}f_1=g_0\,\partial_{z_2}f_1=0\right\}}{\left\{\left(
\begin{array}{c}
  \nabla f_1 \wedge \left(
\begin{array}{c}
  g_1 \\
  g_2 \\
  g_3 \\
\end{array}
\right) \\
  0 \\
\end{array}
\right)\ \Big/\ \mathbf{g}\in A^3\right\}} \\
    & \simeq & \frac{\left\{\mathbf{g}\in A^3\ / \ \nabla f_1\cdot \mathbf{g}=0\right\}}
    {\left\{\nabla f_1\wedge \mathbf{g}\ /\ \mathbf{g}\in A^3\right\}}\oplus\{g\in A\ /\
    g\,\partial_{z_3}f_1=g\,\partial_{z_1}f_1=g\,\partial_{z_2}f_1=0\}. \\
\end{array}$\\

\noindent Le paragraphe suivant va permettre d'expliciter ces
espaces.\\

\subsection{\textsf{Calculs explicites dans le cas particulier où $f_1$ est à variables séparées}}

\noindent Dans ce paragraphe, on considère le polynôme
$f_1=a_1z_1^i+a_2z_2^j+a_3z_3^k$, avec $2\leq i\leq j\leq k$ et $a_j\in \mathbb{C}^*$.\\

\noindent Les dérivées partielles de $f_1$ sont
$\partial_1f_1=ia_1z_1^{i-1}$, $\partial_2f_1=ja_2z_2^{j-1}$ et
$\partial_3f_1=ka_3z_3^{k-1}$.\\

\noindent $\bullet$ On a déjà
\begin{center}
\fbox{$H^0=\mathbb{C}[z_1,\,z_2,\,z_3]/\langle
a_1z_1^i+a_2z_2^j+a_3z_3^k\rangle$}.\\
\end{center}

\noindent $\bullet$ De plus, comme $f_1$ est quasi-homogène, la
formule d'Euler donne
$\displaystyle{\frac{1}{i}z_1\partial_1f_1+\frac{1}{j}z_2\partial_2f_1+\frac{1}{k}z_3\partial_3f_1=f_1}$.
Ainsi, on a l'inclusion $\langle f_1\rangle\subset\langle
\partial_1f_1,\,\partial_2f_1,\,\partial_3f_1\rangle$, donc $$\frac{A}
{\langle
\partial_1f_1,\,\partial_2f_1,\,\partial_3f_1\rangle_A}\simeq
\frac{\mathbb{C}[z_1,\,z_2,\,z_3]}{\langle
\partial_1f_1,\,\partial_2f_1,\,\partial_3f_1\rangle}\simeq
Vect\left(z_1^pz_2^qz_3^r\ /\ p\in\llbracket 0,\,i-2\rrbracket,\
q\in\llbracket 0,\,j-2\rrbracket,\
r\in\llbracket 0,\,k-2\rrbracket\right).$$\\
Enfin, comme $\partial_1f_1$ et $f_1$ sont premiers entre eux, si
$g\in A$ vérifie $g\partial_1f_1=0 \mod \langle f_1\rangle$, alors
$g\in
\langle f_1\rangle$, ie~$g$ est nul dans $A$.\\

\noindent $\bullet$ On détermine maintenant l'ensemble
$\left\{\mathbf{g}=\left(
\begin{array}{c}
  g_1 \\
  g_2 \\
  g_3 \\
\end{array}
\right)\in A^3\ /\ \mathbf{g}\cdot\nabla f_1=0\right\}$ :\\
On a d'abord $\langle
f_1,\,\partial_1f_1,\,\partial_2f_1\rangle=\langle
a_1z_1^i+a_2z_2^j+a_3z_2^k,\,z_1^{i-1},\,z_2^{j-1}\rangle=\langle
z_1^{i-1},\,z_2^{j-1}z_3^k\rangle$. Les seuls monômes qui ne sont
pas dans cet idéal sont donc les éléments $z_1^pz_2^qz_3^r$ avec
$p\in\llbracket
0,\,i-2\rrbracket$, $q\in\llbracket 0,\,j-2\rrbracket$ et $r\in\llbracket 0,\,k-1\rrbracket$.\\
Ainsi tout polynôme $P\in \mathbb{C}[\mathbf{z}]$ s'écrit sous la
forme
$$\displaystyle{P=\alpha f_1+\beta
\partial_1f_1+\gamma
\partial_2f_1+\sum_{\substack{p=0\dots i-2 \\ q=0\dots
j-2 \\ r=0\dots
k-1}}a_{pqr}z_1^pz_2^qz_3^r}.$$\\
Les polynômes $P\in \mathbb{C}[\mathbf{z}]$ tels que
$P\partial_3f_1\in \langle
f_1,\,\partial_1f_1,\,\partial_2f_1\rangle$ sont donc les éléments
$$\displaystyle{P=\alpha f_1+\beta
\partial_1f_1+\gamma
\partial_2f_1+\sum_{\substack{p=0\dots i-2 \\ q=0\dots
j-2 \\ r=1\dots k-1}}a_{pqr}z_1^pz_2^qz_3^r}.$$
On a ainsi calculé $Ann_{\langle f_1,\,\partial_1f_1,\,\partial_2f_1\rangle}(\partial_3f_1)$.\\
L'équation
\begin{equation}\label{31eq1}
\mathbf{g}\cdot\nabla f_1=0 \mod \langle f_1\rangle
\end{equation}
entraîne $g_3\in Ann_{\langle
f_1,\,\partial_1f_1,\,\partial_2f_1\rangle}(\partial_3f_1)$, ie
\begin{equation}\label{31eq3}
g_3=\alpha f_1+\beta\partial_1f_1+\gamma
\partial_2f_1+\sum_{\substack{p=0\dots i-2 \\ q=0\dots
j-2 \\ r=1\dots k-1}}a_{pqr}z_1^pz_2^qz_3^r,
\end{equation}
avec $(\alpha,\,\beta,\,\gamma)\in \mathbb{C}[\mathbf{z}]^3$.\\
D'où
\begin{equation}\label{31eq4}
g_2\partial_2f_1+\gamma\partial_2f_1\partial_3f_1+\sum_{\substack{p=0\dots
i-2 \\ q=0\dots j-2 \\ r=1\dots k-1}}a_{pqr}z_1^pz_2^qz_3^r
\partial_3f_1\in \langle f_1,\,\partial_1f_1\rangle.
\end{equation}
Donc d'après la formule d'Euler,
\begin{equation}\label{31eq5}
\partial_2f_1\left(g_2+\gamma\partial_3f_1-\frac{k}{j}\sum_{\substack{p=0\dots
i-2 \\ q=0\dots j-2 \\ r=1\dots
k-1}}a_{pqr}z_1^pz_2^{q+1}z_3^{r-1}\right)\in \langle
f_1,\,\partial_1f_1\rangle.
\end{equation}
Comme $Ann_{\langle
f_1,\,\partial_1f_1\rangle}(\partial_2f_1)=\langle
f_1,\,\partial_1f_1\rangle$, cette équation équivaut à
$$g_2=-\gamma\partial_3f_1+\frac{k}{j}\sum_{\substack{p=0\dots
i-2 \\ q=0\dots j-2 \\ r=1\dots
k-1}}a_{pqr}z_1^pz_2^{q+1}z_3^{r-1}+\delta
f_1+\varepsilon\partial_1f_1,$$ avec $\delta,\,\varepsilon\in
\mathbb{C}[\mathbf{z}]$. Il s'ensuit que
\begin{equation}\label{31eq6}
g_1\partial_1f_1+\beta\partial_1f_1\partial_3f_1+\varepsilon\partial_1f_1\partial_2f_1+\sum_{\substack{p=0\dots
i-2 \\ q=0\dots j-2 \\ r=1\dots k-1}}a_{pqr}z_1^pz_2^qz_3^r
\partial_3f_1+\frac{k}{j}\sum_{\substack{p=0\dots
i-2 \\ q=0\dots j-2 \\ r=1\dots
k-1}}a_{pqr}z_1^pz_2^{q+1}z_3^{r-1}
\partial_3f_1\in \langle f_1\rangle.
\end{equation}
Et, d'après la formule d'Euler,
\begin{equation}\label{31eq7}
\partial_1f_1\left(g_1+\beta\partial_3f_1+\varepsilon\partial_2f_1-\frac{k}{i}\sum_{\substack{p=0\dots
i-2 \\ q=0\dots j-2 \\ r=1\dots
k-1}}a_{pqr}z_1^{p+1}z_2^qz_3^{r-1}\right)\in \langle f_1\rangle.
\end{equation}
Mais $f_1$ et $\partial_1f_1$ sont premiers entre eux, donc
\begin{equation}\label{31eq8}
g_1=-\beta\partial_3f_1-\varepsilon\partial_2f_1+\frac{k}{i}\sum_{\substack{p=0\dots
i-2 \\ q=0\dots j-2 \\ r=1\dots
k-1}}a_{pqr}z_1^{p+1}z_2^qz_3^{r-1}+\eta f_1,
\end{equation}
avec $\eta\in \mathbb{C}[\mathbf{z}]$.\\

\noindent Finalement
\begin{scriptsize}
$$\left\{\mathbf{g}\in A^3\ /\
\mathbf{g}\cdot\nabla f_1=0 \right\}=\left\{ \left(%
\begin{array}{c}
  \eta \\
  \delta \\
  \alpha \\
\end{array}%
\right)f_1+\nabla f_1\wedge\left(%
\begin{array}{c}
  -\gamma \\
  \beta \\
   -\varepsilon \\
\end{array}%
\right)+\sum_{\substack{p=0\dots i-2 \\ q=0\dots j-2 \\ r=1\dots
k-1}}a_{pqr}z_1^pz_2^qz_3^{r-1}\left(%
\begin{array}{c}
  \frac{k}{i}z_1 \\
  \frac{k}{j}z_2 \\
  z_3 \\
\end{array}%
\right)\ \Big/\
(\alpha,\,\beta,\,\gamma,\,\delta,\,\varepsilon,\,\eta)\in A^6\
\textrm{et}\ a_{pqr}\in \mathbb{C}\right\}.$$
\end{scriptsize}
On en déduit directement les espaces de cohomologie d'indices
impairs
:\\

\begin{center}
\fbox{$\begin{array}{rcl}
\forall\ p\geq 1,\ H^{2p+1} & \simeq & \mathbb{C}^{(i-1)(j-1)(k-1)} \\
 H^{1} & \simeq &  \nabla f_1\wedge\left(\mathbb{C}[\mathbf{z}]/\langle f_1 \rangle\right)^3\oplus\mathbb{C}^{(i-1)(j-1)(k-1)}.\\
\end{array}$}\\
\end{center}

\noindent \\ \underline{Remarque :}\\
On a aussi $\nabla f_1\wedge\left(\mathbb{C}[\mathbf{z}]/\langle
f_1 \rangle\right)^3\simeq \left(\mathbb{C}[\mathbf{z}]/\langle
f_1 \rangle\right)^3\ /\ \{\mathbf{g}\ /\ \nabla f_1\wedge
\mathbf{g}=0\}=\left(\mathbb{C}[\mathbf{z}]/\langle f_1
\rangle\right)^3\ /\ (\mathbb{C}[\mathbf{z}]/\langle f_1
\rangle)\nabla f_1$.\\
De plus, l'application $$\begin{array}{rcl}
  (\mathbb{C}[\mathbf{z}]/\langle f_1\rangle)^2 & \rightarrow & \nabla f_1\wedge\left(\mathbb{C}[\mathbf{z}]/\langle f_1 \rangle\right)^3 \\
  \left(%
\begin{array}{c}
  g_1 \\
  g_2 \\
\end{array}%
\right) & \mapsto & \nabla f_1\wedge\left(%
\begin{array}{c}
  g_1 \\
  g_2 \\
  0 \\
\end{array}%
\right) \\
\end{array}$$ est injective, donc \fbox{$\nabla f_1\wedge\left(\mathbb{C}[\mathbf{z}]/\langle f_1
\rangle\right)^3$ est de dimension infinie}.\\

\noindent \\$\bullet$ Il reste à déterminer l'ensemble
$\left\{\mathbf{g}=\left(
\begin{array}{c}
  g_1 \\
  g_2 \\
  g_3 \\
\end{array}
\right)\in A^3\ /\ \nabla f_1\wedge\mathbf{g}=0\right\}$ :\\
Soit $\mathbf{g}\in A^3$ tel que $\nabla f_1\wedge\mathbf{g}=0$.
Cela signifie que, modulo $\langle f_1\rangle$, $\mathbf{g}$
vérifie le système $\left\{\begin{array}{rcl}
  \partial_2f_1\,g_3-\partial_3f_1\,g_2 & = & 0 \\
  \partial_3f_1\,g_1-\partial_1f_1\,g_3 & = & 0 \\
  \partial_1f_1\,g_2-\partial_2f_1\,g_1 & = & 0 \\
\end{array}\right.$\\
La première équation donne, modulo $\langle
f_1,\,\partial_2f_1\rangle$, $\partial_3f_1\,g_2=0$.\\
Or
$Ann_{\langle f_1,\,\partial_2f_1\rangle}(\partial_3f_1)=\langle
f_1,\,\partial_2f_1\rangle$, donc $g_2=0$,
donc $g_2=\alpha f_1+\beta \partial_2f_1$.\\
Donc
\begin{center}
$\partial_2f_1(g_3-\beta\partial_3f_1)=0\mod \langle f_1\rangle$,
\end{center}
ie $g_3=\gamma f_1+\beta\partial_3f_1$.\\
Enfin, on obtient
\begin{center}
$\partial_3f_1(g_1-\beta\partial_1f_1)=0\mod \langle f_1\rangle$,
\end{center}
ie $g_1=\delta f_1+\beta\partial_1f_1$.\\
Ainsi, $\{\mathbf{g}\in A^3\ /\ \nabla
f_1\wedge\mathbf{g}=0\}=\left\{f_1\left(%
\begin{array}{c}
  \delta \\
  \alpha \\
  \gamma \\
\end{array}%
\right)+\beta\nabla f_1\ /\ \alpha,\,\beta,\,\gamma,\,\delta\in
A\right\}$.\\

\noindent On en déduit les espaces de cohomologie d'indices pairs
:

\begin{center}
\fbox{$\begin{array}{rcl} \forall\ p\geq 2,\ H^{2p} & \simeq & A\
/\ \langle
\partial_1f_1,\,\partial_2f_1,\,\partial_3f_1\rangle\simeq
\mathbb{C}[\mathbf{z}]\ /\ \langle
z_1^{i-1},\,z_2^{j-1},\,z_3^{k-1}\rangle  \\
 & \simeq & Vect\left(z_1^pz_2^qz_3^r\ /\ p\in\llbracket 0,\,i-2\rrbracket,\ q\in\llbracket0,\,j-2\rrbracket,\
 r\in\llbracket0,\,
k-2\rrbracket\right) \simeq \mathbb{C}^{(i-1)(j-1)(k-1)} \\
H^{2} & \simeq &  \{\beta\,\nabla f_1\ /\ \beta\in
A\}\oplus\mathbb{C}^{(i-1)(j-1)(k-1)}\simeq
\mathbb{C}[\mathbf{z}]\ /\ \langle
a_1z_1^i+a_2z_2^j+a_3z_3^k\rangle\oplus\mathbb{C}^{(i-1)(j-1)(k-1)}.\\
\end{array}$}\\
\end{center}

\begin{Rq}$\\$
On obtient en particulier la cohomologie pour les cas où
$f_1=z_1^2+z_2^2+z_3^{k+1}$, $f_1=z_1^2+z_2^3+z_3^4$ et
$f_1=z_1^2+z_2^3+z_3^5$, cas qui correspondent respectivement aux
types $A_k$, $E_6$ et $E_8$ des surfaces de Klein.\\
\end{Rq}

\noindent Le tableau suivant résume les résultats pour
ces trois cas particuliers :\\

\begin{tabular}{|l||l|l|l|l|l|}
  \hline
    & $H^0$ & $H^1$  & $H^2$ & $H^{2p}$ & $H^{2p+1}$ \\
  \hline\hline
  $A_k$ & $\mathbb{C}[\mathbf{z}]\ /\ \langle z_1^2+z_2^2+z_3^{k+1}\rangle$ & $\nabla f_1\wedge\left(\mathbb{C}[\mathbf{z}]\ /\ \langle f_1\rangle\right)^3\oplus \mathbb{C}^k$ & $\mathbb{C}[\mathbf{z}]\ /\ \langle z_1^2+z_2^2+z_3^{k+1}\rangle\oplus \mathbb{C}^k$ &  $\mathbb{C}^k$ & $\mathbb{C}^k$
  \\ \hline
  $E_6$ & $\mathbb{C}[\mathbf{z}]\ /\ \langle z_1^2+z_2^3+z_3^4\rangle$ & $\nabla f_1\wedge\left(\mathbb{C}[\mathbf{z}]\ /\ \langle f_1\rangle\right)^3\oplus \mathbb{C}^6$ & $\mathbb{C}[\mathbf{z}]\ /\ \langle z_1^2+z_2^3+z_3^4\rangle\oplus \mathbb{C}^6$ &  $\mathbb{C}^6$ & $\mathbb{C}^6$
  \\ \hline
  $E_8$ & $\mathbb{C}[\mathbf{z}]\ /\ \langle z_1^2+z_2^3+z_3^5\rangle$ & $\nabla f_1\wedge\left(\mathbb{C}[\mathbf{z}]\ /\ \langle f_1\rangle\right)^3\oplus \mathbb{C}^8$ & $\mathbb{C}[\mathbf{z}]\ /\ \langle z_1^2+z_2^3+z_3^5\rangle\oplus \mathbb{C}^8$ &  $\mathbb{C}^8$ & $\mathbb{C}^8$ \\
  \hline
\end{tabular}\\

\noindent \\ Les cas où $f_1=z_1^2+z_2^2z_3+z_3^{k-1}$ et
$f_1=z_1^2+z_2^3+z_2z_3^3$, ie respectivement $D_k$ et $E_7$
sont étudiés dans le paragraphe suivant.\\

\subsection{\textsf{Calculs explicites pour $D_k$ et $E_7$}}

\subsubsection{\textsf{Cas de $f_1=z_1^2+z_2^2z_3+z_3^{k-1}$, ie $D_k$}}

\noindent Dans ce paragraphe, on considère le polynôme
$f_1=z_1^2+z_2^2z_3+z_3^{k-1}$.\\
Les dérivées partielles de $f_1$ sont $\partial_1f_1=2z_1$,
$\partial_2f_1=2z_2z_3$ et
$\partial_3f_1=z_2^2+(k-1)z_3^{k-2}$.\\

\noindent $\bullet$ On a déjà
$$\fbox{$H^0=\mathbb{C}[\mathbf{z}]/\langle
z_1^2+z_2^2z_3+z_3^{k-1}\rangle.$}$$\\

\noindent $\bullet$ De plus, comme $f_1$ est quasi-homogène, la
formule d'Euler donne
\begin{equation}\label{euler3Dk}
 \displaystyle{\frac{k-1}{2}z_1\,\partial_1f_1+\frac{k-2}{2}z_2\,\partial_2f_1+z_3\,\partial_3f_1=(k-1)f_1}.
\end{equation}
Ainsi, on a l'inclusion $\langle f_1\rangle\subset\langle
\partial_1f_1,\,\partial_2f_1,\,\partial_3f_1\rangle$.\\
De plus, une base de Gröbner de $\langle
\partial_1f_1,\,\partial_2f_1,\,\partial_3f_1\rangle$ est $[z_3^{k-1},\ z_2z_3,\ z_2^2+(k-1)z_3^{k-2},\
z_1]$, donc
$$\frac{A}
{\langle
\partial_1f_1,\,\partial_2f_1,\,\partial_3f_1\rangle_A}\simeq
\frac{\mathbb{C}[z_1,\,z_2,\,z_3]}{\langle
\partial_1f_1,\,\partial_2f_1,\,\partial_3f_1\rangle}\simeq
Vect\left(z_2,\,1,\,z_3,\dots,\,z_3^{k-2}\right).$$\\
Enfin, comme $\partial_1f_1$ et $f_1$ sont premiers entre eux, si
$g\in A$ vérifie $g\partial_1f_1=0 \mod \langle f_1\rangle$, alors
$g\in
\langle f_1\rangle$, ie~$g$ est nul dans $A$, donc $\{g\in A\ /\ g\,\partial_{z_3}f_1=g\,\partial_{z_1}f_1=g\,\partial_{z_2}f_1=0\}=0$.\\

\noindent $\bullet$ On détermine maintenant l'ensemble
$\left\{\mathbf{g}=\left(
\begin{array}{c}
  g_1 \\
  g_2 \\
  g_3 \\
\end{array}
\right)\in A^3\ /\ \mathbf{g}\cdot\nabla f_1=0\right\}$ :\\
Une base de Gröbner de $\langle
f_1,\,\partial_1f_1,\,\partial_3f_1\rangle$ est $[z_1,\,
z_3^{k-1},\ z_2^2+(k-1)z_3^{k-2}]$, donc une base de
$\mathbb{C}[\mathbf{z}]\ /\ \langle
f_1,\,\partial_1f_1,\,\partial_3f_1\rangle$ est $\{z_2^iz_3^j\ /\
i\in\{0,\,1\},\ j\in\llbracket
0,\,k-2\rrbracket\}$.\\
On a déjà résolu l'équation $p\,\partial_2f_1$ dans cet espace ;
sa solution est
$\displaystyle{p=a_{0,k-2}z_3^{k-2}+\sum_{j=0}^{k-2}a_{1,j}z_2z_3^j}$.\\
L'équation
\begin{equation}\label{31Deq1}
\mathbf{g}\cdot\nabla f_1=0 \mod \langle f_1\rangle
\end{equation}
entraîne
\begin{equation}\label{31Deq2}
g_2\partial_2f_1=0 \mod \langle
f_1,\,\partial_1f_1,\,\partial_3f_1\rangle,
\end{equation}
d'où
\begin{equation}\label{31Deq3}
g_2=\alpha f_1+\beta\partial_1f_1+\gamma
\partial_3f_1+az_3^{k-2}+\sum_{j=0}^{k-2}b_jz_2z_3^j,
\end{equation}
avec $(\alpha,\,\beta,\,\gamma)\in \mathbb{C}[\mathbf{z}]^3$.\\
Et
\begin{equation}\label{31Deq4}
g_3\partial_3f_1+\gamma\partial_3f_1\partial_2f_1+az_3^{k-2}\partial_2f_1+\sum_{j=0}^{k-2}b_jz_2z_3^j\partial_2f_1
\in \langle f_1,\,\partial_1f_1\rangle.
\end{equation}
 Or d'après la formule d'Euler (\ref{euler3Dk}) et l'égalité
\begin{equation}\label{31Deq4bis}
 z_3^{k-1}z_2=\frac{1}{2-k}\left(z_2f_1-z_2z_3\partial_3f_1-\frac{1}{2}z_2z_1\partial_1f_1\right)=-\frac{1}{2-k}z_2z_3\partial_3f_1 \mod \langle f_1,\,\partial_1f_1\rangle,
\end{equation}
l'équation (\ref{31Deq4}) devient
\begin{equation}\label{31Deq5}
\partial_3f_1\left(g_3+\gamma\partial_2f_1-\frac{2a}{2-k}z_2z_3-\sum_{j=0}^{k-2}b_j\frac{2}{2-k}z_3^{j+1}\right)\in \langle
f_1,\,\partial_1f_1\rangle.
\end{equation}
Comme $Ann_{\langle
f_1,\,\partial_1f_1\rangle}(\partial_3f_1)=\langle
f_1,\,\partial_1f_1\rangle$, cette équation équivaut à
$$g_3=-\gamma\partial_2f_1+\frac{2a}{2-k}z_2z_3+\sum_{j=0}^{k-2}b_j\frac{2}{2-k}z_3^{j+1}+\delta
f_1+\varepsilon\partial_1f_1,$$ avec
$\delta,\,\varepsilon\in \mathbb{C}[\mathbf{z}]$.\\
On trouve
\begin{equation}\label{31Deq8}
g_1=-\beta\partial_2f_1-\varepsilon\partial_3f_1+\sum_{j=0}^{k-2}b_j\frac{k-1}{k-2}z_1+\frac{a}{2-k}z_2z_1+\eta
f_1,
\end{equation}
avec $\eta\in \mathbb{C}[\mathbf{z}]$.\\

\noindent Finalement, on a
\begin{scriptsize}
$$\left\{\mathbf{g}\in A^3\ /\
\mathbf{g}\cdot\nabla f_1=0 \right\}=\left\{ \left(%
\begin{array}{c}
  \eta \\
  \alpha \\
  \delta \\
\end{array}%
\right)f_1+\nabla f_1\wedge\left(%
\begin{array}{c}
  \gamma \\
  \varepsilon \\
   -\beta \\
\end{array}%
\right)+\sum_{j=0}^{k-2}b_j\left(%
\begin{array}{c}
  \frac{k-1}{k-2}z_1z_3^j \\
  z_2z_3^j \\
  -\frac{2}{2-k}z_3^{j+1} \\
\end{array}%
\right)+a\left(%
\begin{array}{c}
  \frac{1}{2-k}z_2z_1 \\
  z_3^{k-2} \\
  \frac{2a}{2-k}z_2z_3 \\
\end{array}%
\right)\ \Big/\
(\alpha,\,\beta,\,\gamma,\,\delta,\,\varepsilon,\,\eta)\in A^6\
\textrm{et}\ a,\,b_j\in \mathbb{C}\right\},$$
\end{scriptsize}
ainsi que les espaces de cohomologie d'indices impairs
:\\

\begin{center}
\fbox{$\begin{array}{rcl}
\forall\ p\geq 1,\ H^{2p+1} & \simeq & \mathbb{C}^k \\
 H^{1} & \simeq &  \nabla f_1\wedge\left(\mathbb{C}[\mathbf{z}]/\langle f_1 \rangle\right)^3\oplus\mathbb{C}^k.\\
\end{array}$}\\
\end{center}

\noindent \\$\bullet$ Pour montrer que $\{\mathbf{g}\in A^3\ /\
\nabla f_1\wedge\mathbf{g}=0\}=\left\{f_1\,\mathbf{g}+\beta\nabla
f_1\ /\ \mathbf{g}\in
A^3,\ \beta\in A\right\}$, on procède comme dans le cas des variables séparées.\\
On en déduit, en utilisant le premier point $\bullet$, les espaces
de cohomologie d'indices pairs :

\begin{center}
\fbox{$\begin{array}{rcl} \forall\ p\geq 2,\ H^{2p} & \simeq & A\
/\ \langle
\partial_1f_1,\,\partial_2f_1,\,\partial_3f_1\rangle\simeq
Vect\left(z_2,\,1,\,z_3,\dots,\,z_3^{k-2}\right) \simeq \mathbb{C}^k \\
H^{2} & \simeq &  \{\beta\,\nabla f_1\ /\ \beta\in
A\}\oplus\mathbb{C}^k\simeq \mathbb{C}[\mathbf{z}]\ /\ \langle
z_1^2+z_2^2z_3+z_3^{k-1}\rangle\oplus\mathbb{C}^k.\\
\end{array}$}\\
\end{center}

\subsubsection{\textsf{Cas de $f_1=z_1^2+z_2^3+z_2z_3^3$, ie $E_7$}}

\noindent Ici, on a $\partial_1f_1=2z_1,\
\partial_2f_1=3z_2^2+z_3^3$ et $\partial_3f_1=3z_2z_3^2$.\\
La méthode de démonstration est analogue à celle des cas
précédents.\\
Une base de Gröbner de $\langle
\partial_1f_1,\,\partial_2f_1,\,\partial_3f_1\rangle$ est $[z_3^5,\ z_2z_3^2,\ 3z_2^2+z_3^3,\
z_1]$.\\
De même, une base de Gröbner de $\langle
f_1,\,\partial_1f_1,\,\partial_2f_1\rangle$ est $[z_3^6,\
z_2z_3^3,\ 3z_2^2+z_3^3,\ z_1]$.\\
On obtient les résultats suivants :\\

\begin{center}
\fbox{$\begin{array}{rcl}
\forall\ p\geq 1,\ H^{2p+1} & \simeq & \mathbb{C}^7 \\
 H^{1} & \simeq &  \nabla f_1\wedge\left(\mathbb{C}[\mathbf{z}]/\langle f_1 \rangle\right)^3\oplus\mathbb{C}^7.\\
\end{array}$}\\
\end{center}

\begin{center}
\fbox{$\begin{array}{rcl}
 H^0 & = & \mathbb{C}[\mathbf{z}]\ /\ \langle z_1^2+z_2^3+z_2z_3^3\rangle \\
\forall\ p\geq 2,\ H^{2p} & \simeq & A\ /\ \langle
\partial_1f_1,\,\partial_2f_1,\,\partial_3f_1\rangle\simeq
Vect\left(z_2,\,z_2^2,\,1,\,z_3,\,z_3^2,\,z_3^3,\,z_3^4\right) \simeq \mathbb{C}^7 \\
H^{2} & \simeq &  \{\beta\,\nabla f_1\ /\ \beta\in
A\}\oplus\mathbb{C}^k\simeq \mathbb{C}[\mathbf{z}]\ /\ \langle
z_1^2+z_2^3+z_2z_3^3\rangle\oplus\mathbb{C}^7.\\
\end{array}$}\\
\end{center}

\begin{Rq}$\\$
Dans tous les cas étudiés précédemment, il existe un triplet
$(i,\,j,\,k)$ tel que $\{i,\,j,\,k\}=\{1,\,2,\,3\}$, et tel que
l'application
$$\begin{array}{rcl}
  \mathbb{C}[\mathbf{z}]\ /\ \langle \partial_1 f_1,\,\partial_2 f_1,\,\partial_3 f_1
  \rangle & \rightarrow & \{\textrm{Solutions\ dans\ }\mathbb{C}[\mathbf{z}]\ /\ \langle f_1,\,
  \partial_j f_1,\,\partial_k f_1 \rangle\textrm{\ de\ l'équation\ }g\,\partial_i f_1=0 \} \\
  P & \mapsto & z_i\,P \mod \langle f_1,\,\partial_j f_1,\,\partial_k f_1 \rangle \\
\end{array}$$
soit un isomorphisme d'espaces vectoriels.
\end{Rq}

\end{document}